\documentclass{article}

\usepackage{arxiv}

\usepackage[utf8]{inputenc} 
\usepackage[T1]{fontenc}    
\usepackage{hyperref}       
\usepackage{url}            
\usepackage{booktabs}       
\usepackage{amsfonts}       
\usepackage{nicefrac}       
\usepackage{microtype}      
\usepackage{lipsum}
\usepackage{amsmath}
\usepackage{authblk}
\usepackage{comment}
\usepackage{graphicx}
\usepackage{tocbibind}
\usepackage[usenames]{color}
\usepackage{here}

\title{Fluctuation in background synaptic activity controls synaptic plasticity}

\author[1,+]{Yuto Takeda}
\author[2,3,4,5,+]{Katsuhiko Hata \thanks {khata@kokushikan.ac.jp}}
\author[1,+]{Tokio Yamasaki}
\author[6,7]{Masaki Kaneko}
\author[3]{Osamu Yokoi}
\author[3,5]{Chengta Tsai}
\author[1]{Kazuo Umemura}
\author[2]{Tetsuro Nikuni}

\affil[1]{Department of Physics, Faculty of Science Division II, Tokyo University of Science, Tokyo, Japan}
\affil[2]{Department of Physics, Faculty of Science Division I, Tokyo University of Science, Tokyo, Japan}
\affil[3]{Department of Neuroscience, Research Center for Mathematical Medicine, Tokyo, Japan}
\affil[4]{Department of Sports and Medical Science, Kokushikan University, Tokyo, Japan}
\affil[5]{Graduate School of Emergency  Medical System, Kokushikan University, Tokyo, Japan}
\affil[6]{KYB medical service Co., LTD., Tokyo, Japan}
\affil[7]{The Institute of Physical Education, Kokushikan University, Tokyo, Japan}
\affil[+]{These authors contributed equally to this work.}

\date{}
\begin{document}
\maketitle
\begin{abstract}
Synaptic plasticity is vital for learning and memory in the brain. It consists of long-term potentiation (LTP) and long-term depression (LTD). Spike frequency is one of the major components of synaptic plasticity in the brain, a noisy environment. Recently, we mathematically analysed the frequency-dependent synaptic plasticity (FDP) in vivo and found that LTP is more likely to occur with an increase in the frequency of background synaptic activity. Previous studies suggest fluctuation in the amplitude of background synaptic activity. However, little is understood about the relationship between synaptic plasticity and the fluctuation in the background synaptic activity. To address this issue, we performed numerical simulations of a calcium-based synapse model. Then, we found attenuation of the tendency to become LTD due to an increase in the fluctuation of background synaptic activity, leading to an enhancement of synaptic weight. Our result suggests that the fluctuation affect synaptic plasticity in the brain.
\end{abstract}

\thispagestyle{empty}
\newpage

\section{Introduction}
Synaptic plasticity is essential for information processing, including learning and memory. It consists of long term amplification and attenuation of the efficacy in synaptic transmission; long-term potentiation (LTP) and long-term depression (LTD) \cite{RN3}. It has been thought that two mechanisms are involved in the synaptic plasticity. The first is a spike timing of the following presynapse and postsynapse. This spike-timing dependent plasticity (STDP) has been examined in numerous experimental and theoretical studies so far \cite{RN65, RN50, RN102}. LTP is induced by presynaptic firing followed by postsynaptic firing occurring no more than tens of milliseconds, whereas postsynaptic firing preceding presynaptic spikes produces LTD \cite{RN65, RN50, RN51, RN53, RN69, RN70}. The second is a spike frequency \cite{RN18, RN58}. High-frequency inputs in presynaptic neurons induce LTP, whereas the low-frequency firing induces LTD \cite{RN62, RN63, RN64}. This phenomenon is called frequency-dependent synaptic plasticity (FDP) and was formulated as a Bienenstock, Cooper, and Munro (BCM) rule \cite{RN18}.

It has been reported that, in some cases, FDP may be more suitable than STDP for synaptic plasticity in vivo. The in vivo whole-cell patch-clamp recordings in rats revealed that a large amount of endogenous noise in the cerebral cortex limits strict spike timing, essential for STDP \cite{RN75}. A theoretical study also showed that changes in firing rates alone could sufficiently induce synaptic plasticity \cite{RN13}. Moreover, some studies on STDP-based spiking neural networks considering dendritic and axonal propagation delays have suggested the importance of firing frequency \cite{RN104, RN105}.

Recently, we theoretically examined the FDP with in vivo conditions and demonstrated that the output of synaptic plasticity in neurons receiving even the same spike frequency varies by other multifaceted factors, that is, temporal spike patterns, calcium decay time constants and background activities \cite{RN110}. In particular, we showed that a rise in the frequency of background synaptic activity increases the probability of LTP, suggesting the importance of background synaptic activity on synaptic plasticity. This result qualitatively consists with previous experimental studies \cite{RN92, RN93}. Meanwhile, the amplitude of the background synaptic activity usually fluctuates \cite{RN37}. However, little is understood about the relation between the fluctuation in the background event and synaptic plasticity. In this study, we addressed this problem and verified the relation between the coefficient variation for the background noise amplitude and the synaptic plasticity by numerically analyzing a calcium-based model, one of the most suitable models for experimental results \cite{RN1, RN3}. The findings in our study may provide a better understanding of synaptic plasticity in neurons in vivo exposed to noisy background activity.

\section{Materials and Methods}
\label{Method}
\subsection{Model}
We used a model for the FDP based on the calcium control hypothesis by Shouval et al \cite{RN1}. In this model, the postsynaptic calcium concentration affects the time derivative of the synaptic weight.

The dynamics of the synaptic weight $W(t)$ is given as a function of the postsynaptic calcium concentration $Ca(t)$ as follows:

\begin{align}
\frac{d}{dt}W(t) = \eta(Ca(t))[\Omega(Ca(t))-W(t)] \ \ ,\label{eq:W'(t)}
\end{align}

\noindent where $\eta(Ca(t))$ and $\Omega(Ca(t))$ are governed by

\begin{align}
\eta(Ca(t)) &= \left[\frac{p1}{p2+(Ca(t))^{p3}}+p4\right]^{-1} \ \ ,\label{eq:eta(ca)}\\
\Omega(Ca(t)) &= 1+4 \mathrm{sig}(Ca(t)-\alpha_2, \beta_2)- \mathrm{sig}(Ca(t)-\alpha_1, \beta_1), \ \ \label{eq:Omega(ca)} \\
\quad \mathrm{sig}(x, \beta) &:= \exp(\beta x)/[1+\exp(\beta x)]. \ \ \label{eq:sig}
\end{align}

The dynamics of the postsynaptic calcium concentration is described as follows:

\begin{align}
\frac{d}{dt}Ca(t) &= I_{\scalebox{0.5}{NMDA}}(t)-\frac{1}{\tau_{ca}}Ca(t) \label{eq:ca1} \ \ , \\
I_{\scalebox{0.5}{NMDA}}(t, V) &= H(V) \left[I_f \Theta(t) e^{(-t/\tau_f)} + I_s \Theta(t) e^{(-t/\tau_s)}\right] \label{eq:INMDA} \ \ , \\
H(V) &= -P_0 \ G_{\scalebox{0.5}{NMDA}} \frac{(V-V_r)}{1+(Mg/3.57)\exp(-0.062 V)}  \label{eq:H(V)} \ \ .
\end{align}

\noindent
Here, $\tau_{ca}$ indicates the calcium decay time constant. In this study, we fixed $\tau_{Ca}=80$ ms, which is known as representative values in pyramidal cells in the deep cortex (layers V to VI) \cite{RN23, RN5}. The calcium current via the NMDA receptor ($I_{\scalebox{0.5}{NMDA}}$) in Eqs. (\ref{eq:ca1}) and (\ref{eq:INMDA}) is expressed as a function of time ($t$) and postsynaptic membrane potential ($V$). $\Theta(t)$ in Eq. (\ref{eq:INMDA}) is the Heaviside step function.

 The postsynaptic membrane potential is given by
\begin{align}
V(t) &= V_{\rm rest}+V_{\rm epsp}(t)+V_{\rm bg}(t) \ \ \label{eq:V(t)}, \\
V_{\rm epsp}(t) &= \sum_{i} \Theta(t-t_i) \left[e^{-(t-t_i)/\tau_1}-e^{-(t-t_i)/\tau_2}\right] \ \ \label{eq:EPSP(t)},
\end{align}
\noindent
where $V_{\rm rest}$ is the resting membrane potential and $V_{\rm epsp}$ is a depolarization term by EPSPs generated by binding glutamate to the postsynaptic AMPA receptors. $V_{\rm bg}$ is an applied membrane potential by the background synaptic activity, and described as follows:

\begin{eqnarray}
V_{\rm bg}(t) = s \sum_{i} \Theta(t-t_i) \xi_{\rm cv} \left[e^{-(t-t_i)/\tau_1}-e^{-(t-t_i)/\tau_2}\right] \ \ \label{eq:V_{bg}(t)},
\end{eqnarray}
\noindent
where $\xi_{\rm cv}$ follows the normal distribution with average $\langle \xi_{\rm cv} \rangle = 1$ and the coefficient of variation $CV=\langle (\xi_{\rm cv}- \langle \xi_{\rm cv} \rangle)^2 \rangle / \langle \xi_{\rm cv} \rangle$. The $\{ t_i \}$ indicates Poisson processes with 1 Hz mean frequency generated for each simulation. We performed numerical simulations for three cases with different value of $CV=1$, 3 and 5.

All model parameters in this study are listed below \cite{RN1}: \\
$p1=0.1$ s, $p2=p1/10^{-4}$, $p3=3$, $p4=1$ s, $\alpha_1=0.35 \ \mu \mathrm{mol/dm^3}$, $\alpha_2=0.55 \ \mu \mathrm{mol/dm^3}$ and $\beta_1=\beta_2=80 \ \mu \mathrm{mol/dm^3}$, $I_f=0.75$, $I_s=0.25$, $\tau_f=50$ ms, and $\tau_s=200$ ms, $P_0=0.5$, $G_{\scalebox{0.5}{NMDA}}=-1/140 \ \mu \mathrm{mol \cdot dm^{-3}/(m \cdot mV)}$, $Mg=1$, $Vr=130 \ \mathrm{mV}$, $V_{\rm rest}=-65 \ \mathrm{mV}$, $s=20 \ \mathrm{mV}$, $\tau_1 = 50$ ms and $\tau_2 = 5$ ms. 

\subsection{Numerical simulations}
In all numerical simulations, Wolfram Mathematica software version 12.0 was used. For each input frequency, we obtained the time evolution of the postsynaptic calcium concentration $Ca(t)$ and the synaptic weight $W(t)$ by numerically solving Eqs. (\ref{eq:W'(t)}) - (\ref{eq:V_{bg}(t)}). The inter-spike intervals for all inputs are constant. To examine steady states of the system, we calculated the average of the calcium level or the synaptic weight between $8.5 \times 10^4 $ ms to $9.0 \times 10^4 $ ms. Data are expressed as the mean of ten independent experiments and the standard error of the mean (SEM).

\section{Results}
Previously, we demonstrated that the increase in the frequency of background synaptic activities induces the acceleration of the increased rate of the postsynaptic calcium level and the enhancement of synaptic weight \cite{RN110}. The present study aims to investigate the effect of fluctuation in background synaptic activity on the FDP. We used a biophysical model by Shouval et al., in which the change of the synaptic weight is determined by the postsynaptic calcium concentration \cite{RN1, RN2}. This model explains to a large extent experimental data of STDP and FDP obtained so far \cite{RN109}. Thus, we applied presynaptic inputs ranging from 1 to 20 Hz and postsynaptic background activity fixed at 1 Hz to the synapse represented by this model. We examined how a coefficient of variation (CV: varied from 1 to 5 in this study) for an amplitude of the background synaptic activity ($V_{\rm bg}$ in Eq. \ref{eq:V_{bg}(t)}) regulates the frequency-dependence of postsynaptic calcium concentration and synaptic weight.

\begin{figure}[htbp]
\includegraphics[width=\linewidth]{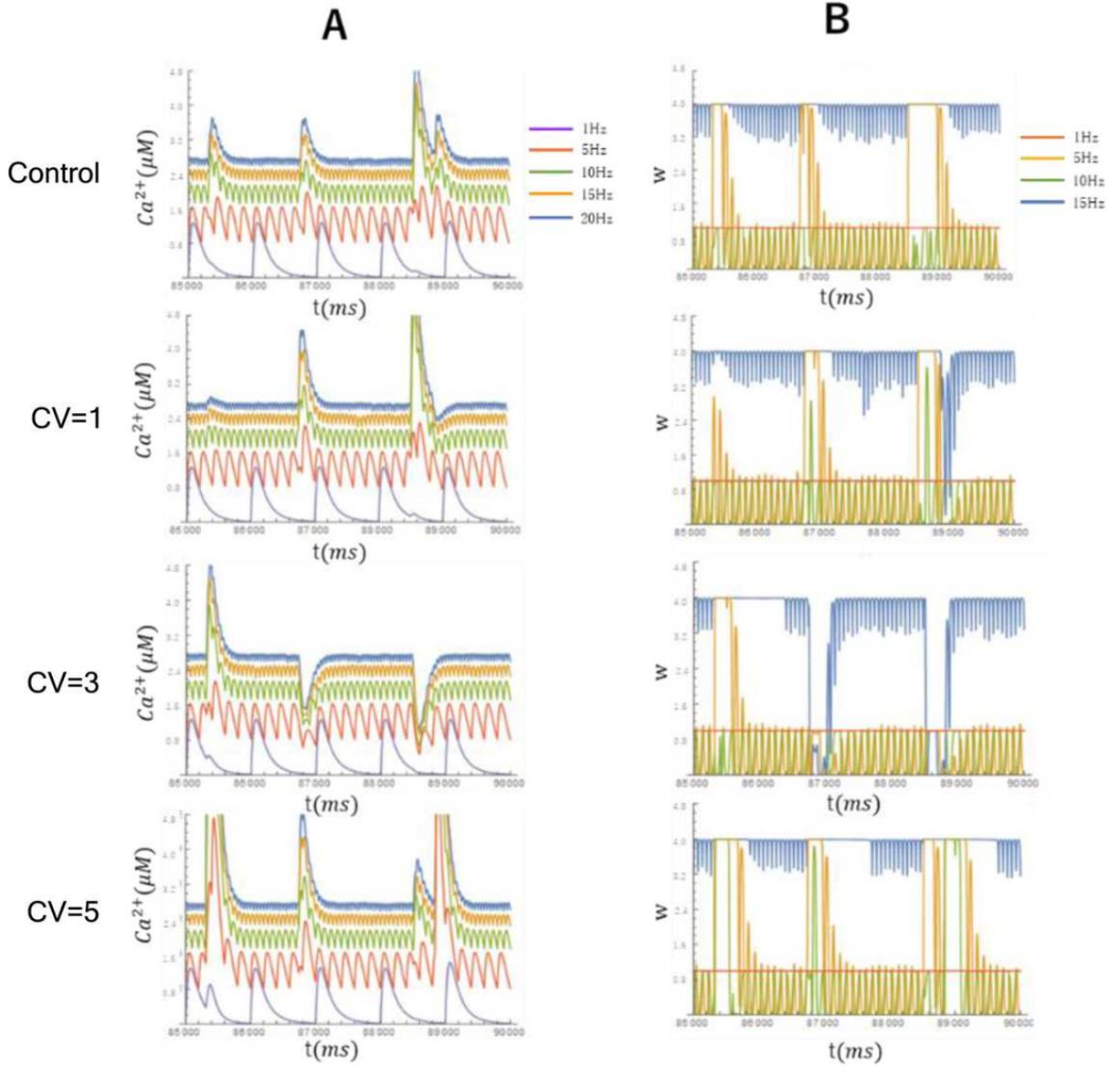}
\caption{Typical simulation results, showing time evolution of the  postsynaptic calcium concentration ($Ca^{2+}$ in A) and synaptic weight ($W$ in B). We evaluated the case that the CV of the postsynaptic background activity ($V_{\rm bg}$), whose frequency is fixed at 1Hz, were 1, 3 and 5. The data without the applied fluctuation is shown as a control. The graphs of the cases where the input frequencies are 1, 5, 10, 15 and 20 Hz are superimposed.
}
 \label{fig1}
\end{figure}

We show representative time evolutions of the postsynaptic calcium concentration ($Ca^{2+}$ in Fig. \ref{fig1}A) and synaptic weight ($W$ in Fig. \ref{fig1}B). Simulations were performed for cases where the CV of $V_{\rm bg}$ were 1, 3 and 5, and where the input frequencies were 1, 5, 10, 15 and 20 Hz. Since we focused on the long-term behaviour of $Ca^{2+}$ and $W$, we calculated the average of the calcium level or the synaptic efficacy between $8.5 \times 10^4 $ ms to $9.0 \times 10^4 $ ms, which is sufficient for the system to reach a steady-state.

\subsection{Relationship between the frequency-dependence of postsynaptic calcium concentration and the CV of an amplitude for the background synaptic activity}
\begin{figure}[htbp]
\includegraphics[width=\linewidth]{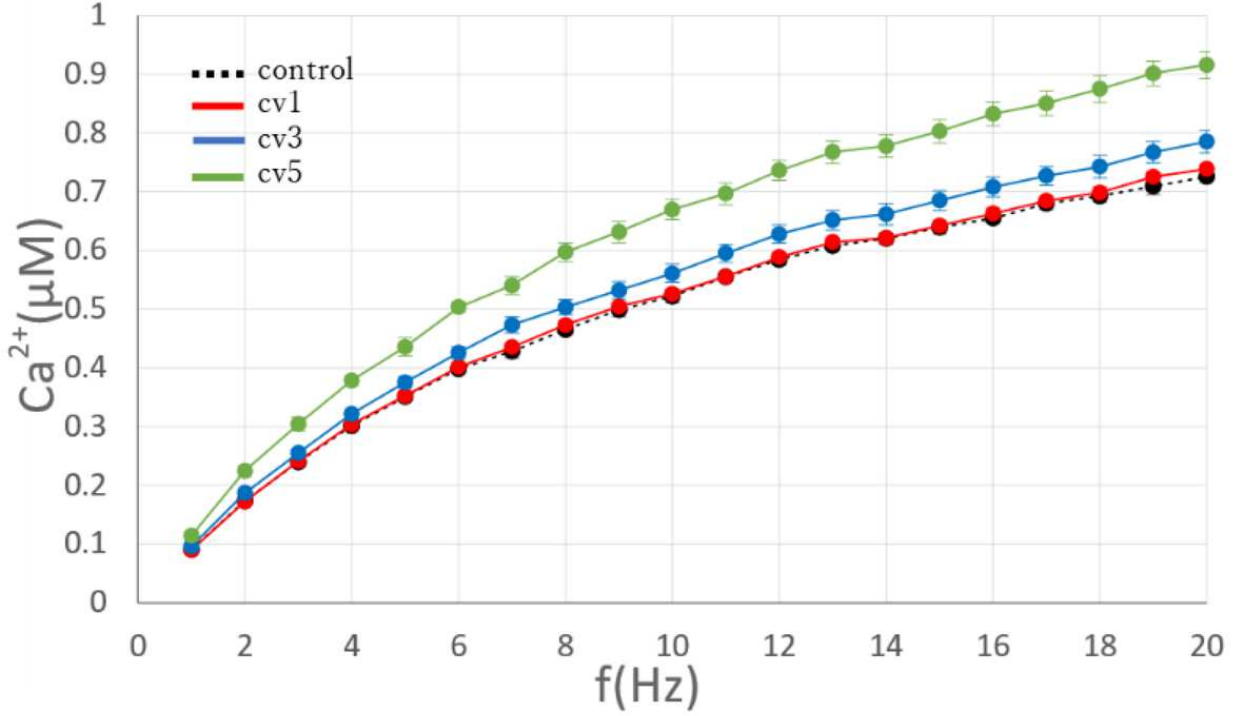}
\caption{The CV of the amplitude for the background synaptic activity regulates the frequency-dependence of the postsynaptic calcium concentration. The x axis shows the input frequency, and the y axis indicates the postsynapitc calcium concentration. A black dotted line indicates the control without applied the fluctuation. The red, blue and green lines show the results of background synaptic activity with fluctuations in the $CV=1$, $3$ and $5$. Error bars indicate the standard error of the mean (SEM).
}
 \label{fig2}
\end{figure}

First, we examined the influence of the increase in the CV of amplitude for the background synaptic activity whose frequency was fixed at 1 Hz on the postsynaptic calcium level. As shown in Fig. \ref{fig2}, the larger the CV, the greater the rate of increase in the postsynaptic calcium concentration. Thus, the rate up of the postsynaptic calcium concentration by increasing input frequency grew with an increase in the background synaptic activity fluctuation.

\subsection{Relationship between the frequency-dependency of synaptic weight and the CV of an amplitude for the background synaptic activity}
\begin{figure}[htbp]
\includegraphics[width=\linewidth]{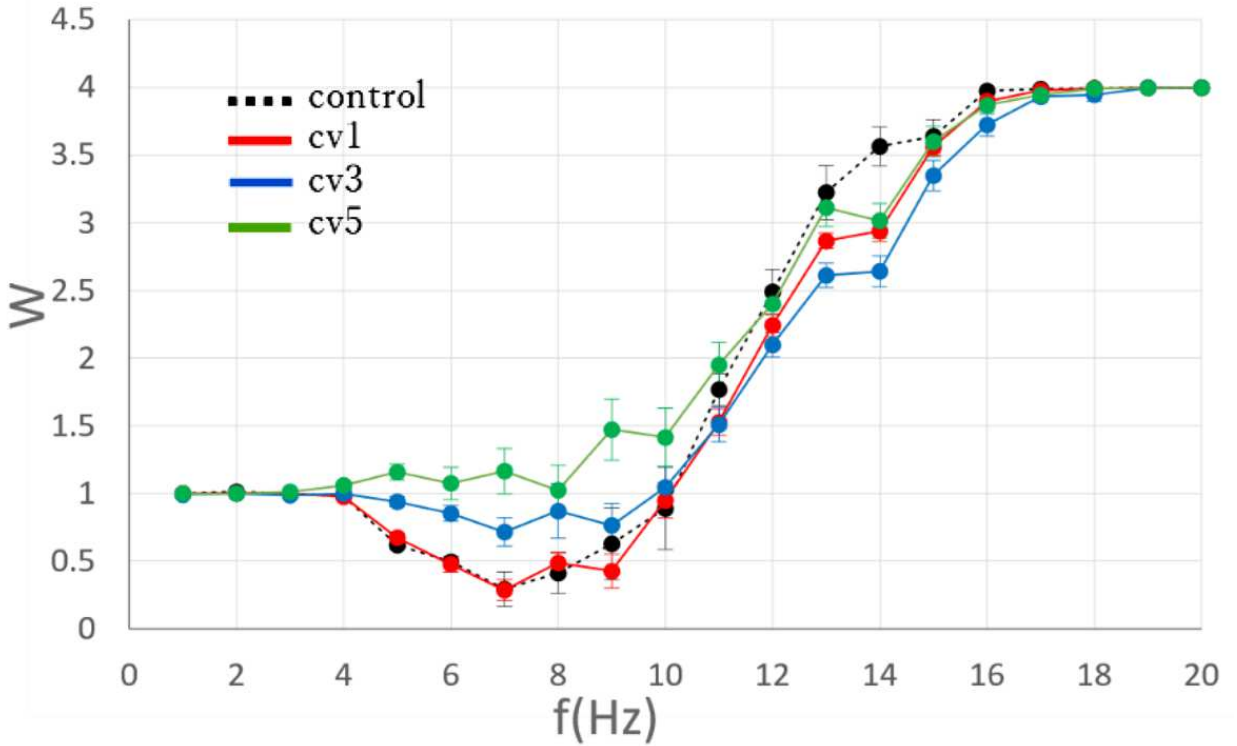}
\caption{
The CV of the amplitude for the background synaptic activity regulates the frequency-dependence of the synaptic strength. The x axis shows the input frequency, and the y axis indicates the relative synaptic weight. $W<1$, $W>1$ and $W=1$ indicates the synaptic strength weakens, becomes more robust and do not change, respectively. A black dotted line indicates the control without applied the fluctuation. In contrast, the results of background synaptic activity with fluctuations in the $CV=1$, $3$ and $5$ are shown by red, blue and green lines. Error bars indicate the standard error of the mean (SEM).
}
 \label{fig3}
\end{figure}

Next, we examined the influence the increase in the CV of amplitude for the background synaptic activity on the synaptic weight. As before, we set the frequency of the background synaptic activity at 1 Hz. The numerical results are plotted in Fig. \ref{fig3}.

\begin{figure}[htbp]
\includegraphics[width=\linewidth]{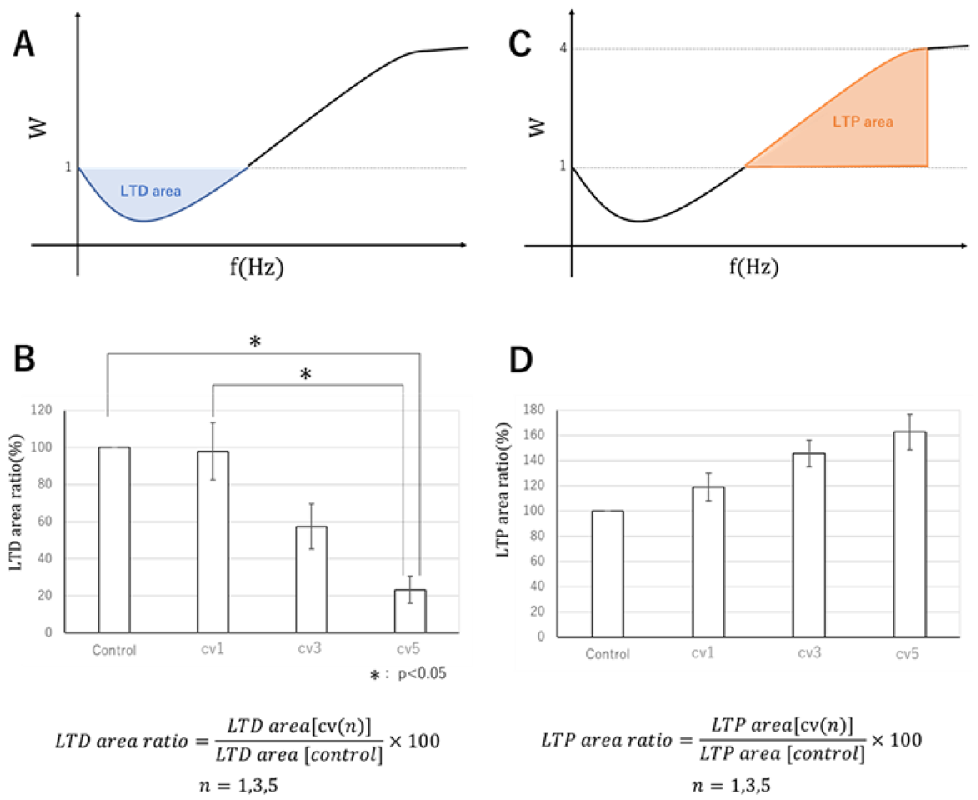}
\caption{
Quantitative data on the change in the synaptic weight by receiving fluctuated background synaptic activity. (A) A schema of LTD-area. The blue part indicates the LTD-area, whose acreage is obtained by Eq. (\ref{eq:LTDarea}). In (B) and (D), control is without the fluctuation of background synaptic activity, while $CV1$, $CV3$ and $CV5$ each indicate the results of fluctuations in background synaptic activity with the corresponding CV. We defined ``LTD-area ratio'' or ``LTP-area ratio'' as the ratio of the LTD- or LTP- area with the fluctuation in the background synaptic activity and without the fluctuation (see formulas in B and D). The bar graph shows the LTD-area ratio (\%) or LTP-area ratio (\%) of the control and synapses receiving fluctuations of the corresponding CV. (B) Relation between the amplitude's fluctuation of background synaptic activity and tendency to become LTD. (C) A schema of LTP-area. The red part indicates the LTP-area, whose acreage is obtained by Eq. (\ref{eq:LTParea}). (D) Relation between the amplitude's fluctuation of background synaptic activity and tendency to become LTP.
}
 \label{fig4}
\end{figure}

To evaluate quantitatively the tendency to become LTP or LTD by changing the CV of the background synaptic activity, we defined ``LTD-area'' and ``LTP-area''. As indicated at the schemas in Figs. \ref{fig4}A and \ref{fig4}C ,  they are given as:

\begin{align}
LTD \ area &:=\int_{0}^{f_0} W(f) df, \  \label{eq:LTDarea} \\
LTP \ area &:=\int_{f_{0}}^{f_{+}} W(f) df. \ \label{eq:LTParea}
\end{align}

\noindent Here, $f_{0}$ in Eqs. (\ref{eq:LTDarea}) and (\ref{eq:LTParea}) indicates the LTD/LTP threshold, which is defined as the frequency at which the synaptic strength first returns to 1 after falling below 1 when the input frequency increases from 0 Hz. In Eq. (\ref{eq:LTParea}), $f_{+}$ is defined as the smaller value of $20$ Hz and the frequency reaching a plateau. Furthermore, to quantify the effect of the fluctuation in the amplitude for the background synaptic activity on synaptic plasticity comparing the control, we calculated ``LTD-area ratio'' and ``LTP-area ratio'' (which are obtained by equations in Figs. \ref{fig4}B and \ref{fig4}D). Thus, we found that the LTD-area ratio decreased as the fluctuation increased. In particular, when $CV = 5$, the LTD-area ratio became significantly smaller than when control and $CV = 1$. On the other hand, the LTP-area ratio tended to go up as the CV increased, but it was not significant.

Thus, an increase in the fluctuation of the background synaptic activity leads to the enhancement of synaptic efficacy mainly through decreasing in the LTD phase. These results suggest that the FDP output (LTP or LTD) varies depending on the size of the fluctuation for background synaptic activity, even if the mean amplitudes and frequency of it is the same.

\section{Discussion}
The postsynapses in vivo receive intense background activity, which induces fluctuations in their membrane potential. The source of it is attributed to electrical noise, including dendrite action potential and synaptic noise \cite{RN37}. The random opening and closing of ion channels in resting-state results in electrical currents, inducing membrane potential fluctuations without presynaptic inputs \cite{RN111}. The primary source of synaptic noise is spontaneous miniature postsynaptic currents (MPSCs). MPSCs are produced by the spontaneous release of neurotransmitters and observed even in the absence of presynaptic input \cite{RN112}. At a glance, such background activity seems not to play a significant role in synaptic plasticity. Still, several experimental and theoretical studies have demonstrated that an appropriate level of background synaptic activity enhances the neural sensitivity and the ability for signal detection \cite{RN92, RN93, RN113}. Our previous study, indicating that an increase in the frequency of the background synaptic activity is likely to cause LTP, also supports these findings \cite{RN110}. In this study, we focused on fluctuation in background activity. We found that it enhances synaptic connections even if the frequency in both presynaptic input and the background activity does not change, as you can see Fig. \ref{fig4} B showing an increase in fluctuation width of the background activity significantly attenuates the LTD phase. Statistical properties of vesicular release can be considered a factor in the CV of fluctuations in background synaptic activity. Several experimental research found synaptic vesicular release counts follow a binomial distribution \cite{RN114, RN115}. The maximum number of vesicles per synapse that affects the CV of fluctuation in postsynaptic membrane potential is constrained by the number of presynaptic docking sites \cite{RN116}. The number of the docking site is in proportion to the number of clusters of voltage-gated calcium channels in the active zone. Both parameters depend on the developmental stage of animals and synaptic size \cite{RN116, RN117}. Further, the clusters of the calcium channels were individually detected in many kinds of synapses. These findings and our results indicate that the CV of fluctuations in background synaptic activity differs depending on anatomical and physiological features, suggesting that fluctuations in the amplitude of background synaptic activity are involved in the plasticity properties in individual neurons \cite{RN117, RN118}.

\section{Conclusions}
We examined the effect of fluctuations in background synaptic activity on synaptic plasticity by numerically solving a calcium-based model. We found that the LTD phase in FDP decreases significantly with an increase in the fluctuation of the background activity, leading to the strengthening of the synaptic efficacy. Thus, we newly found that the amplitude fluctuation of background synaptic activity has a positive effect on the output of FDP.
\vspace{6pt} 

\newpage

\bibliographystyle{nature}
\bibliography{noisepaper}

\begin{thebibliography}{10}

\bibitem{RN3}
Gerstner, W. and Kistler, W.~M.
\newblock {\em Spiking neuron models : single neurons, populations,
  plasticity}.
\newblock Cambridge University Press, Cambridge, U.K. ; New York,  (2002).

\bibitem{RN65}
Song, S., Miller, K.~D., and Abbott, L.~F.
\newblock {\em Nat Neurosci}{ \bf 3}(9), 919--26 (2000).

\bibitem{RN50}
Bi, G.~Q. and Poo, M.~M.
\newblock {\em J Neurosci}{ \bf 18}(24), 10464--72 (1998).

\bibitem{RN102}
Gerstner, W., Kempter, R., van Hemmen, J.~L., and Wagner, H.
\newblock {\em Nature}{ \bf 383}(6595), 76--81 (1996).

\bibitem{RN51}
Markram, H., Lubke, J., Frotscher, M., and Sakmann, B.
\newblock {\em Science}{ \bf 275}(5297), 213--5 (1997).

\bibitem{RN53}
Debanne, D., Gahwiler, B.~H., and Thompson, S.~M.
\newblock {\em J Physiol}{ \bf 507 ( Pt 1)}, 237--47 (1998).

\bibitem{RN69}
Feldman, D.~E.
\newblock {\em Neuron}{ \bf 27}(1), 45--56 (2000).

\bibitem{RN70}
Zhang, L.~I., Tao, H.~W., Holt, C.~E., Harris, W.~A., and Poo, M.
\newblock {\em Nature}{ \bf 395}(6697), 37--44 (1998).

\bibitem{RN18}
Bienenstock, E.~L., Cooper, L.~N., and Munro, P.~W.
\newblock {\em J Neurosci}{ \bf 2}(1), 32--48 (1982).

\bibitem{RN58}
Bliss, T.~V. and Lomo, T.
\newblock {\em J Physiol}{ \bf 232}(2), 331--56 (1973).

\bibitem{RN62}
Dudek, S.~M. and Bear, M.~F.
\newblock {\em Proc Natl Acad Sci U S A}{ \bf 89}(10), 4363--7 (1992).

\bibitem{RN63}
Mulkey, R.~M. and Malenka, R.~C.
\newblock {\em Neuron}{ \bf 9}(5), 967--75 (1992).

\bibitem{RN64}
Artola, A. and Singer, W.
\newblock {\em Trends Neurosci}{ \bf 16}(11), 480--7 (1993).

\bibitem{RN75}
London, M., Roth, A., Beeren, L., Hausser, M., and Latham, P.~E.
\newblock {\em Nature}{ \bf 466}(7302), 123--7 (2010).

\bibitem{RN13}
Graupner, M., Wallisch, P., and Ostojic, S.
\newblock {\em J Neurosci}{ \bf 36}(44), 11238--11258 (2016).

\bibitem{RN104}
Madadi~Asl, M., Valizadeh, A., and Tass, P.~A.
\newblock {\em Sci Rep}{ \bf 7}, 39682 (2017).

\bibitem{RN105}
Madadi~Asl, M., Valizadeh, A., and Tass, P.~A.
\newblock {\em Sci Rep}{ \bf 8}(1), 12068 (2018).

\bibitem{RN110}
Hata, K., Araki, O., Yokoi, O., Kusakabe, T., Yamamoto, Y., Ito, S., and
  Nikuni, T.
\newblock {\em Sci Rep}{ \bf 10}(1), 13974 (2020).

\bibitem{RN92}
Destexhe, A., Rudolph, M., and Pare, D.
\newblock {\em Nat Rev Neurosci}{ \bf 4}(9), 739--51 (2003).

\bibitem{RN93}
Stacey, W.~C. and Durand, D.~M.
\newblock {\em J Neurophysiol}{ \bf 86}(3), 1104--12 (2001).

\bibitem{RN37}
Faisal, A.~A., Selen, L.~P., and Wolpert, D.~M.
\newblock {\em Nat Rev Neurosci}{ \bf 9}(4), 292--303 (2008).

\bibitem{RN1}
Shouval, H.~Z., Bear, M.~F., and Cooper, L.~N.
\newblock {\em Proc Natl Acad Sci U S A}{ \bf 99}(16), 10831--6 (2002).

\bibitem{RN23}
Ahmed, B., Anderson, J.~C., Douglas, R.~J., Martin, K.~A., and Whitteridge, D.
\newblock {\em Cereb Cortex}{ \bf 8}(5), 462--76 (1998).

\bibitem{RN5}
Liu, Y.~H. and Wang, X.~J.
\newblock {\em J Comput Neurosci}{ \bf 10}(1), 25--45 (2001).

\bibitem{RN2}
Shouval, H.~Z. and Kalantzis, G.
\newblock {\em J Neurophysiol}{ \bf 93}(2), 1069--73 (2005).

\bibitem{RN109}
Graupner, M. and Brunel, N.
\newblock {\em Modeling Synaptic Plasticity in Hippocampus: A Calcium-Based
  Approach},  615--644.
\newblock Springer (2018).

\bibitem{RN111}
White, J.~A., Rubinstein, J.~T., and Kay, A.~R.
\newblock {\em Trends Neurosci}{ \bf 23}(3), 131--7 (2000).

\bibitem{RN112}
Fatt, P. and Katz, B.
\newblock {\em Nature}{ \bf 166}(4223), 597--598 (1950).

\bibitem{RN113}
Lu, L., Jia, Y., Kirunda, J.~B., Xu, Y., Ge, M., Pei, Q., and Yang, L.
\newblock {\em Nonlinear Dynamics}{ \bf 95}(2), 1673--1686 (2019).

\bibitem{RN114}
Malagon, G., Miki, T., Llano, I., Neher, E., and Marty, A.
\newblock {\em J Neurosci}{ \bf 36}(14), 4010--25 (2016).

\bibitem{RN115}
Pulido, C., Trigo, F.~F., Llano, I., and Marty, A.
\newblock {\em Neuron}{ \bf 85}(1), 159--172 (2015).

\bibitem{RN116}
Miki, T., Kaufmann, W.~A., Malagon, G., Gomez, L., Tabuchi, K., Watanabe, M.,
  Shigemoto, R., and Marty, A.
\newblock {\em Proc Natl Acad Sci U S A}{ \bf 114}(26), E5246--E5255 (2017).

\bibitem{RN117}
Nakamura, Y., Harada, H., Kamasawa, N., Matsui, K., Rothman, J.~S., Shigemoto,
  R., Silver, R.~A., DiGregorio, D.~A., and Takahashi, T.
\newblock {\em Neuron}{ \bf 85}(1), 145--158 (2015).

\bibitem{RN118}
Holderith, N., Lorincz, A., Katona, G., Rozsa, B., Kulik, A., Watanabe, M., and
  Nusser, Z.
\newblock {\em Nat Neurosci}{ \bf 15}(7), 988--97 (2012).

\end{thebibliography}

\end{document}